
\magnification=1200
\magnification=1200
\def\pagenumbers{\pageno=1\footline={\hss\tenrm\folio\hss}}
\def\today{\ifcase\month
\or January\or February\or March\or April\or May
\or June\or July\or August\or September\or October
\or November \or December\fi \space\number\day, \number\year}
\def\draft{\headline={ \the\pageno
\hfill File:\jobname, Draft Version:\today}}
\countdef\refno=30
\refno=0
\countdef\sectno=31
\sectno=0
\countdef\chapno=32
\chapno=0
\def\ref{\advance \refno by 1 \ifnum\refno<10 \item{ [\the\refno ]} \else
\item{[\the\refno ]} \fi}
\outer\def\section#1#2\par
           {
           \vskip0pt plus .3\vsize\penalty-250\vskip 0pt plus-.3\vsize
           \bigskip\vskip\parskip
           \noindent\leftline{\rlap{\bf #1}
           \hskip 17pt{\bf #2}}
            \nobreak\smallskip}
\def\abstract{\vfill\eject{\bf Abstract}\smallskip}

\def\hangit#1#2\par{\setbox1=\hbox{#1\enspace}
\hangindent\wd1\hangafter=0\noindent\hskip-\wd1
\hbox{#1\enspace}\ignorespaces#2\par}

\def\Zint{{Z \kern -.45 em Z}}
\def\complex{{\kern .1em {\raise .47ex \hbox {$\scriptscriptstyle |$}}
\kern -.4em {\rm C}}}
\def\real{{\vrule height 1.6ex width 0.05em depth 0ex
\kern -0.06em {\rm R}}}
\def\rational{{\kern .1em {\raise .47ex \hbox{$\scripscriptstyle |$}}
\kern -.35em {\rm Q}}}
\def\natural{{\vrule height 1.6ex width .05em depth 0ex \kern -.35em {\rm N}}}
\def\vide{{{\rm O} \kern -0.7em /}}

\parskip 0.5truecm
\baselineskip=12pt
\catcode`\^^?=9
\newskip\zmineskip \zmineskip=0pt plus0pt minus0pt
\mathchardef\mineMM=20000
\newinsert\footins
\def\footnote#1{\let\minesf\empty 
  \ifhmode\edef\minesf{\spacefactor\the\spacefactor}\/\fi
   #1\minesf\vfootnote{#1}}
\def\vfootnote#1{\insert\footins\bgroup
  \interlinepenalty\interfootnotelinepenalty
  \splittopskip\ht\strutbox 
  \splitmaxdepth\dp\strutbox \floatingpenalty\mineMM
  \leftskip\zmineskip \rightskip\zmineskip
\spaceskip\zmineskip \xspaceskip\zmineskip
 \item{#1}\footstrut\futurelet\next\fominet}
\def\fominet{\ifcat\bgroup\noexpand\next \let\next\fmineminet
  \else\let\next\fminet\fi \next}
\def\fmineminet{\bgroup\aftergroup\minefoot\let\next}
\def\fminet#1{#1\minefoot}
\def\minefoot{\strut\egroup}
\def\footstrut{\vbox to\splittopskip{}}
\skip\footins=\bigskipamount 
\count\footins=1000 
\dimen\footins=8in 
\nopagenumbers
$$\vbox{
\vskip 4.5truecm}$$
\centerline{\bf VECTOR RESONANCES FROM A STRONG ELECTROWEAK SECTOR}
\centerline{\bf AT LINEAR COLLIDERS$\;^{*)}$}
\vskip 2.5truecm
\centerline {R. Casalbuoni$\;^{a,b)}$,
P. Chiappetta $\;^{c)}$,
A. Deandrea$\;^{d)}$,
S. De Curtis$\;^{b)}$,}
\smallskip
\centerline {D. Dominici$\;^{a,b)}$
and R. Gatto$\;^{d)}$}
\vskip 5truecm
\centerline{UGVA-DPT 1993/02-805}
\centerline{hep-ph/9303201}
\centerline{February 1993}
\vskip 2truecm
\noindent
\hrule
*) Partially supported by the Swiss National Foundation
\hfill\break\noindent
a) Dipartimento di Fisica, Univ. di Firenze, I-50125 Firenze, Italy.
\hfill\break\noindent
b) I.N.F.N., Sezione di Firenze, I-50125 Firenze, Italy.
\hfill\break\noindent
c) CPT, CNRS, Luminy Case 907, F-13288, Marseille, France.
\hfill\break\noindent
d) D\'ept. de Phys. Th\'eor., Univ. de Gen\`eve, CH-1211 Gen\`eve 4.
\vfill
\eject
\null
$$\vbox{\vskip 1.5truecm}$$
\centerline
{\bf ABSTRACT}
\vskip 1truecm
\noindent
We explore the usefulness of very energetic
linear $e^+e^-$ colliders in the $TeV$ range in studying an alternative
scheme of electroweak symmetry breaking based on a strong interacting sector.
The calculations are performed within the BESS model which contains
new vector resonances. If the mass $M_V$ of the new boson multiplet lies not
far from the maximum machine energy, or if it is lower,
such a resonant contribution would be quite manifest.
A result of our analysis is that also virtual effects are important.
It appears that annihilation into a fermion pair in such  machines, at the
considered luminosities, would improve only marginally on existing limits
if polarized beams are available and left-right asymmetries are measured.
On the other hand, the process of $W$-pair production by $e^+e^-$
annihilation would allow for sensitive tests of the hypothesized
strong sector,
especially if the $W$ polarizations are  reconstructed from their
decay distributions, and the more so the higher the energy of the machine.
An $e^+e^-$ collider with
c.m. energy $\sqrt{s}=500~GeV$ could improve the limits on the model for
the range $500<M_V(GeV)<1000$ when  $W$ polarization is not reconstructed.
If $W$ polarizations are reconstructed, then the bounds improve for the
entire expected range of $M_V$. These bounds become more stringent for larger
energy of the collider.
We have also studied the detectability of the new resonances through the
fusion subprocesses, but this channel does not seem to be
interesting
even for a collider with a c.m. energy $\sqrt{s}=2~TeV$.
\vfill
\eject
\pagenumbers
\hsize=16true cm
\vsize=23.5true cm
\overfullrule=0pt

\def\eps{\epsilon}
\def\csi{\xi}

\def\gp{{g^\prime}}
\def\gs{{g^{\prime\prime}}}

\def \e{{\rm e}}
\def \rs{\sqrt{s}}
\newcount \nfor

\def \form {\global \advance \nfor by 1 \eqno(1.\the\nfor)}
\noindent
{\bf 1. INTRODUCTION}
\bigskip
Several laboratories are at present engaged in projects of $e^+e^-$ linear
colliders with sufficient energy and luminosity to test for new physics.
Among such activity we mention the work being done at SLAC, at KEK,
at Novosibirsk and Serpukhov, at DESY and Darmstadt, at CERN with the
CLIC project, at various laboratories participating to the TESLA concept,
with several other groups and individuals contributing to such developments.
Much interest has of course been devoted to the physics potentialities
of high energy $e^+e^-$
colliders. Studies have been concentrated on a collider having c.m. energy up
to
500 $GeV$, but at the same time possibilities of c.m. energies of 1 or 2 $TeV$
have also been discussed.
A recent workshop at Munich, Annecy, Hamburg was expecially centered on
$e^+e^-$ at 500 $GeV$ c.m. energy [1], and further developments were
discussed at a Conference in Finland [2]. In this paper we shall describe
in some more detail the calculations we had presented in Hamburg [3]
and Saariselk\"a [4] and extend our previous work to $e^+e^-$ colliders
in the $TeV$ energy range.
\par
The sensitivity of future $e^+e^-$ and $pp$ colliders to the
exploration of the electroweak symmetry breaking mechanism has
been largely discussed, both in the context of the   standard model
or of its supersymmetric extension,
and for alternative schemes  where the Higgs field is
composite and an underlying strong theory is responsible for the electroweak
breaking. In this last scenario the longitudinal components of $W$
and $Z$ become strongly interacting in the $TeV$ region. Possible discovery
of this strong interaction is among the physics prospects of
the future $pp$ colliders, LHC and SSC.
Experiments at $e^+e^-$
linear colliders in the $TeV$ region seem to be complementary,
because the structure of the events is relatively simpler than at hadron
colliders, and therefore offer the possibilities for more detailed studies.
\par
We have considered the sensitivity of $e^+e^-$ linear colliders, for
different options of total center of mass energies and luminosities,
to a model which corresponds to a breaking of the electroweak symmetry
due to a strongly interacting sector: breaking electroweak symmetry
strongly (BESS) [5].
In BESS the electroweak symmetry breaking is obtained via a non-linear
realization and no Higgs particles are present. New gauge bosons
$V$ appear and they could be produced as real resonances
if their mass is below the collider energy.
Because of beamstrahlung and synchrotron radiation, in a high energy collider,
one expects in this case
to see dominant peaks below the maximum c.m. energy even
without having to tune the beam energies.
If instead the masses  of the $V$ bosons are  higher than the maximum c.m.
energy, they would give rise to indirect effects in the
$e^+e^-\rightarrow f^+f^-$ and  $e^+e^-\rightarrow W^+W^-$  cross sections,
which we discuss quantitatively  below.
\par
The BESS model appears as the simplest definite scheme against which to test
for a possible strong electroweak sector.
The vector resonances of the BESS model are bound states of a
strongly interacting sector. In this sense they are similar to ordinary
$\rho$ vector mesons, or to the techni-$\rho$ particle of
technicolor theories.
An effective lagrangian describing in an unified way mass terms and
interactions of the standard electroweak gauge bosons and these new
vector resonances $V$ was derived in ref. [5] as a gauged non linear
$\sigma$-model. The BESS model contains as additional parameters the
mass $M_V$ of the new bosons (which are a degenerate triplet of an additional
$SU(2)$ gauge group), their gauge coupling $\gs$ (which
is assumed to be much larger than $g$, and $\gp$), and a parameter
$b$ specifying the direct coupling of $V$ to the fermions.
The standard model (SM) is recovered in the limit $\gs\to\infty$, and
$b=0$. Mixings of the ordinary gauge bosons to the $V$'s are of the order
$O(g/\gs)$. Due to these mixings, $V$ bosons are coupled to fermions
even for $b=0$. Furthermore these couplings are still present in
the $M_V\to\infty$ limit, and therefore the new gauge boson effects
do not decouple in the large mass limit.
\par
In this paper we will be mainly interested in the study of the effects
due to vector resonances of the strongly interacting
electroweak symmetry breaking sector. We will extend our previous analysis
which was dedicated to a machine in the range of $500~GeV$ [3,4].
Our description of the vector resonances is rather general and after a
convenient specification of the parameters is also apt to describe a standard
techni-$\rho$ ($\rho_T$) state. The production
of $\rho_T$ at $e^+e^-$ machines was analyzed  by M.E. Peskin at
Saariselk\"a [6] (see also references therein). Similar studies were also
done by Iddir et al. [7].
They implement the strongly interacting regime by inserting
in the Born amplitude for $e^+e^-\to W^+ W^-$ strong final state corrections
 from $WW\to WW$. The corrections can be described by the Omn\'es function
which is approximated through a rescaling of the Gounaris-Sakurai
model [8]. These authors note that the most significant sign of a vector
resonance in  $W^+W^-$ would manifest itself in the backward hemisphere, since
the transverse channel $W_TW_T$ dominates forward due to $\nu$-exchange
in $t$-channel.
\par
In his summary talk at Saariselk\"a J.L. Barklow [9]
concludes that a $\rho_T$ state
of $M_{\rho_T}=1.7~TeV$ would be visible at machines with c.m. energies
greater or of the order of 1 $TeV$ already for a luminosity of 50 $fb^{-1}$.
K. Hikasa [10] has proposed to look for possible strong $WW$ scattering
effects by using the interference between $W_LW_L$ and $W_TW_T$ in the
$WW$ pair production process. These would manifest themselves in
correlations between the azimuthal angles of the decay fermions from
$W^+$ and $W^-$, specifically in the dependence of the distribution
 from the sinus of the sum of the two azimuthal angles for $W^+$ and $W^-$
decay (both decaying leptonically, or one leptonically, the other into
$c\bar s$).
\par
The BESS model at LEPII and at a collider energy of $\sqrt{s}=600~GeV$ has
been studied in ref. [11].
\par
Using the recent LEP1 data combined with UA2/CDF data and low energy
electroweak data, we have first derived the present bounds on the BESS
parameter space.
We have studied how these limits can be improved for the following three
options for the future machines: $\sqrt{s}=500$, 1000,
2000 $GeV$, with integrated  luminosity
of 20, 80, 20 $fb^{-1}$ respectively (for the parameters of various
designs see ref. [12]).
\par
We have analyzed
cross-sections and asymmetries for the channel $e^+e^-\rightarrow f^+f^-$
and $e^+e^-\rightarrow W^+W^-$.
For the purposes of our calculation we have also assumed that it will
be possible to separate
$e^+e^-\rightarrow W^+_L W^-_L$, $e^+e^-\rightarrow W^+_L W^-_T$, and
$e^+e^-\rightarrow W^+_T W^-_T$. The distribution of the $W$ decay angle
in its c.m. frame depends indeed in very distinct ways from its helicity,
being peaked forward (backward) with respect to the production direction
for positive (negative) helicity or at $90^o$ for zero helicity.
One may hope that by looking at the different $W$ decay modes it will
be possible  to extract such a  very useful information.
\par
At very energetic linear colliders the pair production of $W^+W^-$ through
$W^+W^-$, $\gamma \gamma$, $\gamma Z$, $ZZ$ fusion have to be taken into
account. Since the final states for the annihilation process, for the fusion
subprocesses through $W^+W^-$, or $\gamma$, $Z$ are different, we treat
them as three independent reactions and examine their sensitivities
to the BESS model parameters.
\par
One clear advantage of $e^+e^-$ collisions with respect to $pp$ is that all
decay modes from a $WW$ or $ZZ$ pair may contribute to the observable
signatures, and not only those which include at least a leptonic mode.
\par
One has to stress that an
important advantage of a linear collider is the possibility one can
envisage of a large longitudinal beam polarization.
\par
We find that the differential cross-sections to $f\bar f$ would
allow to improve the already existing limits on BESS from LEP1 and UA2/CDF
if longitudinally polarized beams will be available, particularly
by considering left-right asymmetries in hadron production.
\par
A considerable improvement of the limits on BESS is obtained
by the study of the differential $WW$ cross section. As a matter of fact
the cross section acquires in BESS a term which rises linearly
with  $s$ plus a constant term which is the most relevant at the
energies considered here, whereas in the SM the cross-section
goes down as $1/s$.
\par
This improvement becomes even more effective if the $W$ polarization
is reconstructed, because the new gauge bosons are mainly coupled
to longitudinal $W$'s.
\par
We will show that, if no deviations from SM
expectations are seen at colliders in the $TeV$ range, the parameter
space left to the BESS model is very small.
Direct search for $V$ bosons at hadronic colliders has been extensively
studied.
In this case the discovery of the neutral resonance $V^0$ looks difficult.
It has however been
shown that LHC and SSC are ideal to pin down a charged resonance up to 2 $TeV$
through the $WZ$ channel. We see that very energetic
$e^+e^-$ linear colliders are complementary to hadron colliders.
\par
$W$-fusion reactions can directly inform on $WW$ scattering. They are expected
at a very energetic $e^+e^-$ collider.
In principle, such a process could be interesting because it would allow
to study a wide range of mass spectrum for the $V$ resonance also if the
energy of the machine is not adjustable downward.
Note however that in such linear colliders beamstrahlung will already
automatically provide for a spectrum of lower initial energies.
Besides, as usual in fusion reactions,
the c.m. energy available for the two colliding $W$'s is strongly lowered
with respect to the original $e^+e^-$ energy. This makes the exploration for
possible effects of a high mass strong electroweak sector via $W$-fusion
less encouraging, also in view of the backgrounds from processes leading
to the same final states and utilizing the full beam energy.
In fact we find that also at a $2~TeV$ $e^+e^-$ energy the number of events
is too low for detectability.
\par
Already in 1983 Ginzburg et al. [13] had suggested the possibility of
obtaining an energetic photon beam by collision of the electron bunches with a
laser beam operating in the visible spectrum.
This technique should allow keeping a high luminosity for the photon beam from
the back-scattered laser and should allow for a photonic spectrum mostly
concentrated at the highest energies,
not much smaller than the electron energy.
Such a behaviour is quite different from that of the expected beamstrahlung
photons concentrated at the lower energies, and, even more so, from that of
the bremstrahlung photons. The possibility would allow for energetic
photon-photon collisions and electron-photon collisions.
For the purpose of the present paper, where we are interested in a possible
strong electroweak sector, $\gamma \gamma$ collisions would appear of interest
if resonant behaviours are present in states of zero angular momentum
which can couple to two real photons. These may be the pseudogoldstone
states as in general expected in these models [14] when one considers
the terms due to the Adler-Bell-Jackiw anomaly. Also interesting will
be the production of pairs or more of such pseudos from $\gamma \gamma$.
As far as the vector-type resonances, such as in the simplest BESS
considered here, they will not be prominent from an initial state of two
real photons. Their contribution will be through rescattering between
$W$'s in $\gamma \gamma$ and in $\gamma e$ collisions.
In $\gamma e$ collisions, contributions will also appear through the
modification of the leptonic vertices which are as prescribed in BESS.
In both cases the additional terms coming from the BESS model
are negligible.
\par
The paper is organized as follows. We will recall in section 2 some basics
of the BESS model and the present limits on parameter space. In section
3 the process $e^+e^- \rightarrow \overline{f}f$ is examined.
The process $e^+e^- \rightarrow W^+_{L,T} W^-_{L,T}$ is discussed in
section 4. Section 5 is devoted to the fusion subprocesses
$e^+e^- \rightarrow W^+_{L,T} W^-_{L,T}e^+e^-$ and
$e^+e^- \rightarrow W^+_{L,T} W^-_{L,T}{\bar \nu}\nu$. We conclude in section
6.
\bigskip
\newcount \nfor

\def \form {\global \advance \nfor by 1 \eqno(2.\the\nfor)}
\noindent
{\bf 2. THE BESS MODEL}
\bigskip
The increasing accumulation of precise measurements of the electroweak
parameters gives the possibility of finding some hints for going beyond the
standard model. Through mixing effects the contribution of vector resonances
 from a possible strong sector would affect masses and
couplings of ordinary gauge bosons. Therefore precise measurements of the width
of $Z$, its mass and forward backward asymmetries, performed at LEP,
allow for restrictions on the unknown parameters of the BESS model
which are: the mass $M_V$, the direct coupling to fermions $b$, and the gauge
coupling constant $\gs$ of the $V$ bosons.
\par
BESS is a non-renormalizable
theory; therefore, when radiative corrections
are considered, a cut-off $\Lambda$ has to be introduced. This cut-off
plays the role of the parameter $m_H$, Higgs mass, of the SM.
The strategy we follow in our analysis is to consider for BESS
the same one-loop radiative corrections as in the SM and to interpret
$m_H$ as the cut-off.
\par
Using the following  LEP1 data averaged on the four LEP experiments,
as communicated at the Dallas Conference [15]:
$$
\eqalign{
&M_Z=91.187\pm 0.007~GeV\cr
&\Gamma_Z=2492\pm 7~MeV\cr
&\Gamma_h=1737.1\pm 6.7~MeV\cr
&\Gamma_{\ell}=83.33\pm 0.30~MeV\cr
&A_{FB}^{\ell}=0.0154\pm 0.0027\cr
&A_{\tau}^{pol}=0.140\pm 0.018\cr
&A_{FB}^{b}=0.098\pm 0.012\cr}
\form
$$
and
$$
{M_W\over M_Z}=0.8807\pm 0.0031
\form
$$
 from CDF/UA2 [16], we can derive
bounds on the BESS model, that we express as $90\%$ C.L.
contours in the plane $(b,g/\gs)$ for given $M_V$ (see Fig. 1).
\par
The comparison with BESS
includes one-loop electroweak radiative corrections calculated
with a cut-off of 1 $TeV$ and for  $\alpha_s=0.12$.
The bounds in Fig. 1 are almost independent of the mass of the new
resonances $V$ and of the choice of the cut-off.
The region gets shifted by changing the values of
$\alpha_s$, $m_{top}$, becoming narrower for increasing $m_{top}$.
So the region of
negative $b$ values is practically excluded for "light" top masses.
\par
By including the low energy data coming
 from cesium parity violation experiments the bounds do not significantly
change. In the context of BESS,
LEP 200 is expected to increase only marginally the sensitivity over LEP1.
The relevant modification will be brought by the more accurate determination
of $M_W$.
\par
In order to compare with a standard techni-$\rho$, one has to take
$b=0$, choose
$$
{g\over\gs}=\sqrt{2}{M_W\over M_{\rho_T}}
\form
$$
and identify $M_{\rho_T}$ with $M_V$ [17].
\bigskip
\newcount \nfor

\def \form {\global \advance \nfor by 1 \eqno(3.\the\nfor)}
\noindent
{\bf 3. SENSITIVITY FROM $e^+e^-\rightarrow f^+f^-$}
\bigskip
In the discussion of high energy linear colliders,
we start with the fermion channels.
Our analysis is
based on the following observables:
$$\eqalign{
&\sigma^{\mu},~~R=\sigma^h/\sigma^{\mu}\cr
&A_{FB}^{e^+e^- \to \mu^+ \mu^-},~~ A_{FB}^{e^+e^- \to {\bar b} b}\cr
&A_{LR}^{e^+e^- \to \mu^+  \mu^-},~~A_{LR}^{e^+e^- \to {\bar b} b},~~
A_{LR}^{e^+e^- \to {had}} \cr
}\form
$$
where
$A_{FB}$ and $A_{LR}$ are the forward-backward and left-right asymmetries,
and $\sigma^{h(\mu)}$ is the total hadronic ($\mu^+\mu^-$) cross section.
Assuming that one can measure the final $W$ polarization, we can add
to our observables the longitudinal and transverse polarized $W$
differential cross sections and asymmetries.
\par
To evaluate these observables within the BESS model, we recall that
in the neutral sector the couplings of the fermions to the gauge bosons
$Z$ and $V$ are [5,18]
$$
e \left (v^f_Z+\gamma_5 a^f_Z\right )\gamma_\mu Z^\mu +
e \left (v^f_V+\gamma_5 a^f_V\right )\gamma_\mu V^\mu
\form
$$
where $v^f_{Z,V}$ and $a^f_{Z,V}$ are the vector and axial-vector
couplings given by
$$
\eqalign{
&v^f_Z={1\over{\sin 2\theta}} \left( A T_3^L+2BQ_{e.m.}\right)\cr
&a^f_Z={1\over{\sin 2\theta}} AT_3^L\cr
&v^f_V={1\over{\sin 2\theta}}\left( CT_3^L+2DQ_{e.m.}\right)\cr
&a^f_V={1\over{\sin 2\theta}}CT_3^L\cr}
\form
$$
where
$$
\eqalign{
&A={{\cos \csi}\over{\cos\psi}}(1+b)^{-1}\Big[1+b\sin^2\theta\big(1-
{{\tan\csi}\over{\tan\theta\sin\psi}}\big)\Big]\cr
&B=-\sin^2\theta{{\cos \csi}\over{\cos\psi}}\Big(1-{{\tan\csi\sin\psi}\over
{\tan\theta}}\Big)\cr
&C={{\sin \csi}\over{\cos\psi}}(1+b)^{-1}\Big[1+b\sin^2\theta\big(1+
{{\cot\csi}\over{\tan\theta\sin\psi}}\big)\Big]\cr
&D=-\sin^2\theta{{\sin \csi}\over{\cos\psi}}\Big(1+{{\cot\csi\sin\psi}\over
{\tan\theta}}\Big)\cr}
\form
$$
with $\sin\theta=\gp/\sqrt{g^2+\gp^2}$ and $e=g\sin\theta\cos\psi$.
The mixing angles in the $M_V>>M_W$ and  large $\gs$ limit are
$$
\eqalign{
&\csi=-{\cos{2\theta}\over{\cos\theta}} {g\over\gs}\cr
&\psi=2\sin\theta{g\over\gs}\cr}
\form
$$
The total cross section for the process $e^+e^-\rightarrow f^+f^-$
is given by (at tree level)
$$
\sigma = {\pi\alpha_{em}^2 s\over 3}\sum_{h_f,h_e}|F(h_f,h_e)|^2
\form
$$
with $\alpha_{em}=e^2/(4\pi)$ and
$$
F(h_f,h_e)=-{1\over s}e_f+{(v^f_Z+h_f a^f_Z)(v_Z+h_e a_Z)
\over {s-M_Z^2+iM_Z\Gamma_Z}}+{(v^f_V+h_f a^f_V)(v_V+h_e a_V)
\over {s-M_V^2+iM_V\Gamma_V}}
\form
$$
where $h_f,~h_e=\pm 1$ are the helicities of $f$ and $e$ respectively, $e_f$
is the electric charge of $f$ (in units of $e_{\rm proton}=1$),
$v_{Z,V}=v^e_{Z,V}$ and $a_{Z,V}=a^e_{Z,V}$, and
$\Gamma_{Z,V}$ are the widths of the neutral gauge bosons.
The partial width of the $V$ bosons corresponding to decays into
fermion-antifermion and $WW$ is given by
$$
\Gamma_V^{h}+3 (\Gamma_V^{l}+\Gamma_V^{\nu})+\Gamma_V^W
\form
$$
where $\Gamma_V^h$ is the sum of all quark-antiquark  widths (including
top), with
$$
\Gamma_V^f={{M_V\alpha_{em}}\over 3}\left ({v^f_V}^2+{a_V^f}^2\right )
\form
$$
$$
\eqalign{
\Gamma_V^W=&{M_V\over 48}\alpha_{em} g^2_{VWW}
\left(1-4{M_W^2\over M_V^2}\right)^{3/2}\left({M_V\over M_W}\right)^4\cr
&\times\left [1+20\left({M_W\over M_V}\right)^2+12\left({M_W\over M_V}\right)^4
\right]\cr}
\form
$$
$$
g_{VWW}={\cos^2\phi\over \tan\theta} \left( {\sin\csi\over \cos\psi}-
\tan\theta\tan\psi\cos\csi\right)+{\cos\csi\sin^2\phi\over 2\sin\theta}
{\gs\over g}
\form
$$
and, in the limit $M_V>>M_W$ and  large  $\gs$
$$
\phi=-{g\over\gs}
\form
$$
The forward-backward asymmetry in the present case is given by
$$
A_{FB}^{e^+e^-\to f^+ f^-}={x\over {1+{1\over 3}x^2}}
{{(1-P)\sum_{h_f,h_e}h_fh_e|F(h_f,h_e)|^2+2P\sum_{h_f}h_f|F(h_f,1)|^2}\over
{(1-P)\sum_{h_f,h_e}|F(h_f,h_e)|^2+2P\sum_{h_f}|F(h_f,1)|^2}}
\form
$$
where $x$ is the detector acceptance ($x\le 1$), and
$P$ is the degree of longitudinal polarization of the
electron beam. In the following analysis we will assume $x=1$.
\par
The left-right asymmetry is given by
$$
A_{LR}^{e^+e^-\to f^+ f^-}=P{{\sum_{h_f,h_e}h_e|F(h_f,h_e)|^2}\over
{\sum_{h_f,h_e}|F(h_f,h_e)|^2}}
\form
$$
The notations are the same as for the forward-backward asymmetry.
\par
In the following numerical analysis, following the existing studies of 500
GeV $e^+e^-$ linear
colliders [3,19], we assume a relative systematic error in luminosity of
${{\delta {\cal L}}/ {\cal L}}=1\%$ and ${{\delta\epsilon_{\rm hadr}}/
\epsilon_{\rm hadr}}=1\%$ (which is perhaps an optimistic choice due to the
problems connected with the $b$-jet reconstruction),
$\delta\epsilon_{\mu}/\epsilon_{\mu}=0.5\%$,
where $\epsilon_{hadr,\mu}$ denote the selection efficiencies. We shall
also assume the same systematic errors for the 1 and 2 $TeV$ machines.
Finally we have considered an integrated luminosity $L=20~fb^{-1}$ for
$\sqrt{s}=500~GeV$, $L=80~fb^{-1}$ for $\sqrt{s}=1~TeV$ and
$L=20~fb^{-1}$ for $\sqrt{s}=2~TeV$. These integrated luminosities correspond
to about one year ($10^7~sec.$) of running. One should of course take
into account beamstrahlung effects. However for two body final states,
as we consider here, the practical effect is a reduction of the luminosity.
This means that with the assumed nominal luminosity one has to run for a
correspondingly longer period.
\par
In the case we cannot reach the mass of the resonance we can get restrictions
on the parameter space by combining the observables of eq. (3.1). Throughout
this paper we assume $m_{top}=150~GeV$ and $\Lambda=1~TeV$. Our results are
shown in Figs. 2 and 3. Unfortunately the most sensitive observables
are the left-right asymmetries, which means that, if the beams are not
polarized, one does practically get no advantage over LEP1 from this channel.
\par
The contours shown in Fig. 2 correspond to the regions which are allowed
at 90\% C.L. in the plane $(b,g/\gs)$, assuming that the BESS deviations
for the observables of eq. (3.1) from the SM predictions are within the
experimental errors. The results are obtained assuming a longitudinal
polarization of the electron $P_e=0.5$ (solid line) and $P_e=0$ (dashed line).
In Fig. 2 we assume $\sqrt{s}=500~GeV$ and $M_V=600~GeV$.
As it is clear there is no big improvement
with respect to the already existing bounds from LEP1. Increasing the
energy of the machine does not drastically change the results.
We have also explored the sensitivity with respect to $M_V$, by choosing
$b=0$ and $P_e=0.5$. For instance, in Fig. 3,
we show the allowed region in the plane $(M_V,g/\gs)$ to be compared with
the already existing limit $g/\gs<0.06$ from LEP1 at $b=0$ (see Fig. 1).
Here we consider both the case of $\sqrt {s}= 500~GeV$ (solid line) and
$\sqrt {s}= 1000~GeV$ (dashed line). The bounds improve only for $M_V$
close to the value of the energy of the machine.
\par
All these conclusions become much more negative if one assumes a higher
systematical error for the hadron selection efficiency.
\bigskip
\newcount \nfor

\def \form {\global \advance \nfor by 1 \eqno(4.\the\nfor)}
\noindent
{\bf 4. SENSITIVITY FROM $e^+e^-\rightarrow W^+W^-$}
\bigskip
In this Section we will consider the $WW$ channel, which is expected to
be more sensitive, at high energy, than the $f\bar f$ channel to effects
coming from a strongly interacting electroweak symmetry breaking sector.
In the case of a vector resonance this is simply due to the strong coupling
between the longitudinal $W$ bosons and the resonance. Furthermore this
interaction, in general, destroys the fine cancellation among the $\gamma-Z$
exchange and the neutrino contribution occurring in the SM. This effect gives
rise, in the case of the BESS model, to a differential cross-section
increasing linearly with $s$. However, by using the explicit
expression given below one can show that the leading term in $s$ is suppressed
by a factor $(g/\gs)^4$. Therefore the effective deviation at the energies
considered here is given only by the constant term, which is of the order
$(g/\gs)^2$.
\par
We shall consider the $WW$ channel, for one $W$  decaying leptonically
and the other hadronically.  The main reason for choosing this decay channel
is to get a clear signal to reconstruct the polarization of the $W$'s
[20]. Let us consider first the following observables:
$$
\eqalign{
&{d\sigma \over {d\cos\theta}}(e^+ e^-\to W^+ W^-)\cr
& A_{LR}^{{ e^+ e^- \to W^+ W^-}}=(
{d\sigma \over {d\cos\theta}}(P_{e}=+P)-
{d\sigma \over {d\cos\theta}}(P_{e}=-P))/
{d\sigma \over {d\cos\theta}}\cr}
\form
$$
where $\theta$ is the $e^+e^-$ center of mass scattering angle.
In the $e^+e^-$ center of mass frame the angular distribution
$d\sigma/d\cos\theta$ and the left-right asymmetry read [11]
$$
\eqalign{
{{d\sigma}\over {d\cos\theta}}=& {{2\pi\alpha_{em}^2 p}\over {\sqrt{s}}}
\Big{\{} 2a_W^4\left [{4\over M^2_W}+p^2\sin^2\theta
\left ( {1\over M_W^4}+{4\over t^2}\right )\right ]\cr
&+G_1p^2\left [ {{4s}\over M^2_W}+
\left ( 3+{{sp^2}\over M^4_W}\right )\sin^2\theta\right ]\cr
&+G_1^{\prime}
\left [ 8\left ( 1+{{M^2_W}\over t}\right )+
16 {{p^2}\over M^2_W}+{{p^2}\over s}\sin^2\theta
\left ( {{s^2}\over M_W^4}-2{s\over M_W^2}-4{s\over t}\right )
\right ] \Big{\}}\cr}
\form
$$
and
$$
\eqalign{
A_{LR}(\cos\theta)=&-P
{{2\pi\alpha_{em}^2 p}\over {\sqrt{s}}}
\Big{\{} 2a_W^4\left [{4\over M^2_W}+p^2\sin^2\theta
\left ( {1\over M_W^4}+{4\over t^2}\right )\right ]\cr
&+G_2p^2\left [ {{4s}\over M^2_W}+
\left ( 3+{{sp^2}\over M^4_W}\right )\sin^2\theta\right ]\cr
&+G_1^{\prime}
\Big [ 8\left ( 1+{{M^2_W}\over t}\right )+
16 {{p^2}\over M^2_W}\cr
&+{{p^2}\over s}\sin^2\theta
\left ( {{s^2}\over M_W^4}-2{s\over M_W^2}-4{s\over t}\right )
\Big ] \Big{\}} {\Big /}{{d\sigma}\over {d\cos\theta}}\cr}
\form
$$
where
$$
\eqalign{
&p={1\over 2}\sqrt{s}(1-4M^2_W/s)^{1/2}\cr
&t=M_W^2-{1\over 2}s[1-\cos\theta (1-4M_W^2/s)^{1/2}]\cr}
\form
$$
The quantity $a_W$ is the $\nu e W$ coupling
$$
a_W={1\over 2 \sqrt{2}\sin\theta} (1+b)^{-1}\left({\cos\phi\over\cos\psi}
-{b\over 2} {\gs\over g}{\sin\phi\over\cos\psi}\right)
\form
$$
and
$$
\eqalign{
G_1&=\left ({{e_e}\over s}\right )^2+
(v_Z^2+a_Z^2)g^2_{ZWW}{1\over {(s-M_Z^2)^2+M_Z^2\Gamma_Z^2}}+
2{e_e\over s}v_Zg_{ZWW} {{s-M_Z^2}\over {(s-M_Z^2)^2+M_Z^2\Gamma_Z^2}}\cr
&+(v_V^2+a_V^2)g^2_{VWW}{1\over {(s-M_V^2)^2+M_V^2\Gamma_V^2}}+
2{e_e\over s}v_Vg_{VWW} {{s-M_V^2}\over {(s-M_V^2)^2+M_V^2\Gamma_V^2}}\cr
&+2 (v_Zv_V+a_Za_V)g_{ZWW}g_{VWW} {{(s-M_Z^2)(s-M_V^2)+M_Z\Gamma_Z M_V\Gamma_V}
\over {[(s-M_Z^2)^2+M_Z^2\Gamma_Z^2][(s-M_V^2)^2+M_V^2\Gamma_V^2]}}\cr}
\form
$$
$$
\eqalign{
G_1^{\prime}&=a^2_W [{{e_e}\over s}\cr
&+g_{ZWW}(v_Z+a_Z)
{{s-M_Z^2}\over {(s-M_Z^2)^2+M_Z^2\Gamma_Z^2}}\cr
&+g_{VWW}(v_V+a_V){{s-M_V^2}\over {(s-M_V^2)^2+M_V^2\Gamma_V^2}}]\cr}
\form
$$
$$
\eqalign{
G_2&=2\Big({e_e\over s}a_Z g_{ZWW}{{s-M_Z^2}\over {(s-M_Z^2)^2+M_Z^2
\Gamma_Z^2}}+
{e_e\over s}a_V g_{VWW}{{s-M_V^2}\over {(s-M_V^2)^2+M_V^2\Gamma_V^2}}\cr
&+a_Zv_Zg_{ZWW}^2{1\over {(s-M_Z^2)^2+M_Z^2\Gamma_Z^2}}
+a_Vv_Vg_{VWW}^2{1\over {(s-M_V^2)^2+M_V^2\Gamma_V^2}}\cr
&+(a_Zv_V+v_Za_V)g_{ZWW}g_{VWW}{{(s-M_Z^2)(s-M_V^2)+M_Z\Gamma_Z M_V\Gamma_V}
\over {[(s-M_Z^2)^2+M_Z^2\Gamma_Z^2][(s-M_V^2)^2+M_V^2\Gamma_V^2)]}}\Big)\cr}
\form
$$
where
$$
g_{ZWW}={\cos^2\phi\over \tan\theta} \left( {\cos\csi\over \cos\psi}+
\tan\theta\tan\psi\sin\csi\right)
\form
$$
and $g_{VWW}$ as given in eq. (3.11).
\par
Assuming that the final $W$ polarization can be reconstructed by using the
$W$ decay distributions,
it is convenient to examine the cross sections for $W_LW_L$, $W_TW_L$, and
$W_TW_T$. One has [21]
$$
\eqalign{
{{d\sigma_{LL}}\over {d\cos\theta}}&= {{2\pi\alpha_{em}^2 p}\over {\sqrt{s}}}
\Big{\{} {{a_W^4}\over 8}{1\over M_W^4}{1\over t^2}
[ s^3 (1+\cos^2\theta)-4M_W^4 (3s+4 M_W^2)\cr
&-4(s+2M_W^2)p\sqrt {s} s \cos\theta  ]\sin^2\theta\cr
&+{1\over {16}}G_1{1\over M_W^4}\sin^2\theta
(s^3-12sM_W^4-16M_W^6)\cr
&+G_1^{\prime}\sin^2\theta {1\over {2t}} [
ps\sqrt{s}\cos\theta{1\over {2M_W^4}}(s+2M_W^2)\cr
&-{1\over {4 M_W^4}}(s^3-12sM_W^4-16M_W^6) ]
\Big{\}}\cr}
\form
$$
$$
\eqalign{
{{d\sigma_{TL}}\over {d\cos\theta}}&= {{2\pi\alpha_{em}^2 p}\over {\sqrt{s}}}
\Big{\{} {{a_W^4}}{1\over {t^2 M_W^2}}
[ s^2 (1+\cos^4\theta)+4M_W^4 (1+\cos^2\theta)\cr
&-4(4p^2+s\cos^2\theta )p\sqrt {s}  \cos\theta
+2s(s-6M_W^2)\cos^2\theta-4sM_W^2 ]\cr
&+2G_1s{p^2\over M_W^2}(1+\cos^2\theta)\cr
&+2G_1^{\prime}
{{p\sqrt{s}}\over {tM_W^2}}[
\cos\theta (4p^2+s\cos^2\theta)-
2p\sqrt{s}(1+\cos^2\theta)]
\Big{\}}\cr}
\form
$$
$$
\eqalign{
{{d\sigma_{TT}}\over {d\cos\theta}}&= {{2\pi\alpha_{em}^2 p}\over {\sqrt{s}}}
\Big{\{} {2{a_W^4}}{1\over {t^2 }}
[ s (1+\cos^2\theta)-2M_W^2-2p\sqrt {s}\cos\theta ]\sin^2\theta\cr
&+2G_1{p^2}\sin^2\theta+G_1^{\prime}
{{\sin^2\theta}\over {2t}}[4p\sqrt{s}\cos\theta -8p^2]
\Big{\}}\cr}
\form
$$
The left-right asymmetries for longitudinal and/or transverse polarized
$W$ can be easily obtained as in eq. (4.3) by substituting $G_1$ by $G_2$
in eqs. (4.10), (4.11), (4.12), and dividing by the corresponding
differential cross section.
In the case of the $WW$ channel we assume ${{\delta B}/ B}=0.005$ [22], where
$B$ denotes the product of the branching ratio for  $W\rightarrow hadrons$
and that for $W\rightarrow leptons$, and we assume $1\%$ for the acceptance.
The assumed value is $B=0.29$.
\par
Figs. 4, 5 show the deviations of the BESS model with respect to the SM
with values of the BESS parameters $b=0$ and $g/\gs=0.05$,
for the two cases $\rs=1~TeV$, $M_V=1.5~TeV$,
and $\rs=2~TeV$, $M_V=2.5~TeV$ respectively. Furthermore we plot both the
unpolarized and the longitudinally polarized differential cross-section
for decay of one $W$ leptonically and the other hadronically.
The systematic errors are the same as
discussed before and the statistical errors are evaluated assuming
integrated luminosities of $80~fb^{-1}$ and $20~fb^{-1}$ for machines
at $\rs= 1~TeV$ and $\rs=2~TeV$ respectively. We observe that, as already
noticed [7, 9], the bigger deviations are away from the forward
region. In the longitudinal channel the deviations are much bigger and
concentrated in the central region.
\par
To discuss the restrictions on the parameter space for masses of the
resonance a little higher than the available energy we have here taken into
account the experimental efficiency. We have assumed
an overall detection efficiency of 10\% including the branching ratio
$B=0.29$ and the loss of luminosity from
beamstrahlung [20].
This gives an effective branching ratio of about 0.1. We have again considered
an integrated luminosity of $20~fb^{-1}$, $80~fb^{-1}$ and $20~fb^{-1}$
for the three cases $\rs=0.5,~1~{\rm and}~ 2~TeV$.
\par
For a collider at
$\rs=500~GeV$ the results are illustrated in Fig. 6. The contours have
been obtained by taking 18 bins in the angular region restricted by
$|\cos\theta|< 0.95$. This figure illustrates the 90\% C.L. allowed regions
for $M_V=600~GeV$
obtained by considering the unpolarized $WW$ differential cross-section
(dotted line), the $W_LW_L$ cross section (dashed line),
and the combination of the left-right asymmetry with all the
differential cross-sections for the different final $W$ polarizations
(solid line). We see that
already at the level of the
unpolarized cross-section we get important restrictions
with respect to LEP1. In Fig. 7 we have examined the possibility that the
total number of events is reduced of a factor 0.5 from the losses due to the
reconstruction of the polarization of the $W$'s. We see that, least to say,
this does not reduce the efficiency of the machine with respect to LEP1
in restricting the BESS parameter space.
\par
For colliders with $\rs=1,~2~TeV$ and for $M_V=1.2~{\rm and}~2.5~TeV$
respectively, the allowed region, combining all the observables,
reduces in practice to a line and the analysis is better discussed in the
plane $(M_V,g/\gs)$, as we shall see later on. Therefore, even the
unpolarized $WW$ differential cross section measurements can improve the
bounds, as shown in Fig. 8 for $\rs= 2~TeV$ and  $M_V=2.5~TeV$.
\par
Finally in Figs. 9-11 we assume $b=0$, that is no direct
coupling of the $V$ to the fermions, and we analyze the allowed region
in the plane $(M_V,g/\gs)$. We work in the same hypotheses as before.
Recalling that the bound on $g/\gs$ from LEP1 is about 0.06 for $b=0$,
we see that if the $W$ polarization is not observed the machine at
$\rs=500~GeV$ improves the LEP1 limit up to $M_V\approx 1~TeV$.  If
the $W$ polarization is measured the LEP1 limit is improved in all $M_V$
range. For colliders at $\rs=1,~2~TeV$ we get a big improvement on
the bound for all the situations.
\bigskip
\newcount \nfor

\def \form {\global \advance \nfor by 1 \eqno(5.\the\nfor)}
\noindent
{\bf 5. SENSITIVITY FROM FUSION SUBPROCESSES}
\bigskip
Another mechanism to produce $W^+ W^- $ pairs is the fusion of a pair
of ordinary gauge bosons, each being initially emitted from an electron or a
positron. In the so called effective-W approximation the initial $W,Z,\gamma$
are assumed to be real and the cross section for producing a  $W^+ W^- $ pair
is obtained by a convolution of the fusion subprocess with the luminosities of
the initial $W,Z,\gamma$ inside the electrons and positrons.
There are two fusion subprocesses which contribute to produce $W^+ W^- $ pairs.
The first one is $e^+e^-
\rightarrow W^+_{L,T} W^-_{L,T} e^+ e^-$. It is mediated by $W^{\pm}$
and $V^{\pm}$ exchanges in the $t$ and $u$ channels.
The second fusion subprocess we consider is $e^+e^-
\rightarrow W^+_{L,T} W^-_{L,T}{\overline \nu} \nu$.
It is mediated by $\gamma, Z$ and $V^{0}$ exchanges in the
$s$ and $t$ channels. Both processes get a contribution from the gauge
boson quadrilinear couplings.
\par
In principle, the fusion processes are interesting because they allow
to study a wide range of mass spectrum for the $V$ resonance from one
given $e^+e^-$ c.m. energy.
\par
In the $e^+e^-$ center-of-mass frame the invariant mass distribution
$d\sigma/dM_{WW}$ reads
$$
\eqalign{
{{d\sigma}\over {dM_{WW}}}&={1\over 4\pi s} {1\over M^2_{WW}}
\sum_{i,j}\sum_{l1,l2}
\int^{M_{WW}^2/4}_{(p_T^2)_{min}}
d{p_T^2}\int^{-\log{\sqrt{\tau}}}_{\log{\sqrt{\tau}}}dy ~f^{l1}_i(\sqrt{\tau}
\e^{y})f^{l2}_j(\sqrt{\tau}\e^{-y})\cr
&\cdot {p'\over p} {1\over\sqrt{M^2_{WW}-4 p_T^2}}
|M({V^{l1}_i V^{l2}_j} \rightarrow W^+_{l3}
W^-_{l4})|^2}
\form
$$
where $p_T$ is the transverse momentum of the outgoing $W$,
$\tau=M_{WW}^2/s$,
$p$ and $p'$ are the absolute values of the three momenta
for incoming and outgoing pairs
of vector bosons: $p=(E_1^2-M_1^2)^{1/2}=(E_2^2-M_2^2)^{1/2}$ and
$p'=(\sqrt{M_{WW}}/2) (1-4 M_W^2/M_{WW}^2)^{1/2}$
with $E_i$ the fraction of the electron
(or positron)
energy of the vector boson $V_i$ with mass $M_i$ and helicity $l_i$.
The structure functions $f$ appearing in the previous formula are
$$
\eqalign{
f^+(x)=&{{\alpha_{em}}\over{4\pi}} {{[(v+a)^2+(1-x)^2(v-a)^2]}\over{x}}
\log{s\over{M^2}}\cr
f^-(x)=&{{\alpha_{em}}\over{4\pi}} {{[(v-a)^2+(1-x)^2(v+a)^2]}\over{x}}
\log{s\over{M^2}}\cr
f^{0}(x)=&{{\alpha_{em}}\over{\pi}} (v^2+a^2) {{1-x}\over{x}}
}
\form
$$
and represent the probability of having inside the electron a vector boson of
mass $M$ with fraction $x$ of the electron energy. In eq. (5.2) $v$ and $a$
are the vector and axial-vector couplings of the gauge bosons to fermions.
More precisely for the photon $v=-1$ and $a=0$, for $Z$ $v=v^f_Z$ and $a=a^f_Z$
as given in eq. (3.3), and for $W$ $v=a=a_W$ as
given in eq. (4.5). The quadrilinear
couplings are
$$
\eqalign{
&g_{WWWW}= {{e^2}\over{\sin^2\theta}} {{\cos^4\phi}\over{\cos^2\psi}}
+ {{\gs^2}\over{4}} \sin^4\phi\cr
&g_{WWZZ}={{e^2}\over{\cos^2 \phi}} g^2_{ZWW}+
{{\gs^2}\over{4}} \sin^2\phi \sin^2\xi \cos^2\psi \cr
&g_{WW\gamma\gamma}= e^2 \cos^2\phi + {{\gs^2}\over{4}} \sin^2\phi \sin^2\psi
\cr
&g_{WWZ\gamma}= e^2 g_{ZWW} - {{\gs^2}\over{4}} \sin^2\phi \sin\xi
\sin\psi \cos\psi}
\form
$$
with $g_{ZWW}$ as given in eq. (4.9) and the mixing angles $\csi$, $\psi$
and $\phi$ in eqs. (3.5) and (3.12).
\par
The amplitudes of the vector bosons scattering processes within the BESS
model are:
$$
\eqalign{
M(W^+_1W^-_2 \rightarrow W^+_3W^-_4)&=-i e^2 {f_s\over s} -i e^2
g^2_{ZWW} {f_s\over {s-M^2_Z}}\cr
& -i e^2 g^2_{VWW} {f_s\over {s-M^2_V+i \Gamma_V M_V}}
+(s\rightarrow t)\cr
& + ig_{WWWW} [2(\eps_1 \cdot \eps_4^*)(\eps_2 \cdot \eps_3^*)
-(\eps_1 \cdot \eps_2) (\eps_3^* \cdot \eps_4^*)\cr
&-(\eps_1 \cdot \eps_3^*)(\eps_2 \cdot \eps_4^*)]\cr}
\form
$$
$$
\eqalign{
M(\gamma_1\gamma_2 \rightarrow W^+_3W^-_4)&=-i e^2 {h_t\over {t-M^2_W}}
+(t\rightarrow u)\cr
& - ig_{WW\gamma\gamma} [2(\eps_1 \cdot \eps_2)(\eps_3^* \cdot \eps_4^*)
-(\eps_1 \cdot \eps_3^*) (\eps_2 \cdot \eps_4^*)\cr
&-(\eps_1 \cdot \eps_4^*)(\eps_2 \cdot \eps_3^*)]\cr}
\form
$$
$$
\eqalign{
M(\gamma_1 Z_2 \rightarrow W^+_3W^-_4)&=-i e^2 g_{ZWW} {h_t\over {t-M^2_W}}
+(t\rightarrow u)\cr
& - ig_{WWZ\gamma} [2(\eps_1 \cdot \eps_2)(\eps_3^* \cdot \eps_4^*)
-(\eps_1 \cdot \eps_3^*) (\eps_2 \cdot \eps_4^*)\cr
&-(\eps_1 \cdot \eps_4^*)
(\eps_2 \cdot \eps_3^*)]\cr}
\form
$$
$$
\eqalign{
M(Z_1 Z_2 \rightarrow W^+_3W^-_4)&= -i e^2
g^2_{ZWW} {h_t\over {t-M^2_W}}\cr
& -i e^2 g^2_{VWW} {h_t\over {t-M^2_V+i
\Gamma_V M_V}}
+(t\rightarrow u)\cr
& - ig_{WWZZ} [2(\eps_1 \cdot \eps_2)(\eps_3^* \cdot \eps_4^*)
-(\eps_1 \cdot \eps_3^*) (\eps_2 \cdot \eps_4^*)\cr
&-(\eps_1 \cdot \eps_4^*)
(\eps_2 \cdot \eps_3^*)]\cr}
\form
$$
where
$$
\eqalign{
f_s=&[(\eps_1 \cdot \eps_2) (p_2-p_1)_{\lambda} -2(\eps_1 \cdot p_2)
\eps_{2 \lambda} + 2 (\eps_2 \cdot p_1) \eps_{1 \lambda} ] \cr
&\cdot [(\eps_3^* \cdot \eps_4^*) (p_3-p_4)^{\lambda} -2(\eps_4^* \cdot p_3)
\eps_{3}^{* \lambda} + 2 (\eps_3^* \cdot p_4) \eps_4^{* \lambda} ]}
\form
$$
and $f_t$ can be deduced from $f_s$ with the substitution $p_2 \leftrightarrow
-p_3$ and $\eps_2 \leftrightarrow \eps^*_3$, while
$$
\eqalign{
h_t=&[2(\eps_1 \cdot p_3) \eps^*_{3 \lambda} - (\eps_1 - \eps^*_3)
(p_3 +p_1)_{\lambda}+2(p_1 \cdot \eps^*_3) \eps_{1 \lambda}] \cr
&\cdot [2(\eps_2 \cdot p_4) \eps^{* \lambda}_4-(\eps_2 \cdot \eps_4^*)
(p_4+p_2)^{\lambda}+2(p_2 \cdot \eps^*_4) \eps^{\lambda}_2]  \cr
&+ (\eps_1 \cdot \eps^*_3) (p_1^2 - p_3^2) (\eps_2 \cdot \eps^*_4)
(p^2_2 - p_4^2)/M_W^2}
\form
$$
and $h_u$ can be deduced from $h_t$ with the substitution $p_3 \leftrightarrow
p_4$ and $\eps_3^* \leftrightarrow \eps^*_4$.
\par
In eqs. (5.4-7) $g_{ZWW}$ and $g_{VWW}$ are given in eq. (4.9) and (3.11)
respectively and $\Gamma_V$ is the width of the $V$ resonance
(we have not explicitly inserted the widths of the standard gauge bosons
because they are irrelevant for this calculation).
\par
For comparison, the amplitudes within the SM
can be obtained from eqs. (5.4-8)
by taking all the trilinear and quadrilinear vector boson couplings in the
limit $\gs\to\infty$ and $b\to 0$ and adding the contribution due to the
Higgs boson exchange (see ref. [23]).
We are not considering here all the non-annihilation graphs
contributing to the processes
$e^+e^-\rightarrow W^+W^-e^+e^-$ and $e^+e^-\rightarrow W^+W^-\nu\bar \nu$
(at the order $\alpha_{em}^4$) in which the final $W$'s are emitted from the
electron (positron) legs. This because we expect their contribution to lie
mostly in a kinematical region different from the one we are interested in
($p_T\sim M_W$).
\par
The result of our analysis is that there are not significative differences
between the SM and the BESS model differential cross section
in the case of the process $e^+e^-
\rightarrow W^+W^-e^+e^-$. This is due, first of all, to the
absence of the $s$ channel
exchange of the $V$ resonance, secondly,
to the dominance of the $\gamma\gamma$ fusion  contribution, and to the
the fact that in the BESS model
the couplings of the
photon to the fermions and to $W^+W^-$ are the  same as in the SM.
\par
Concerning the process
$e^+e^-\rightarrow W^+W^-\nu {\overline \nu}$, we have evaluated the
differential cross sections $d\sigma/d M_{WW}$ both for the SM
with $M_H=100~GeV$ and for the
BESS model. The only channel which turns out to be useful is the one
corresponding
to longitudinally polarized final $W$'s.
The results are illustrated in Figs. 12, 13 for two different choices of
the BESS parameters.
In Fig. 12 we compare $d\sigma/\d M_{WW} (LL)$ for the SM (dashed line) and
for the BESS model (solid line) for $\sqrt {s}=1.5~TeV$, $b=0.01$,
$\gs=13$ and $M_V=1~TeV$.
In Fig. 13 and $\sqrt {s}=2~TeV$, the mass of the resonance is $M_V=1.5~TeV$
and the other
BESS parameters are the same as in Fig. 12.
In both cases we are not applying any cuts except for $(p_T)_{min}=10~GeV$.
For simplicity we have only introduced the partial width of eq. (3.8).
\par
Although theoretically there is a clear difference between the curves
of the two models, the experimental situation is quite different.
\par
Let us first consider the case of $\sqrt{s}=1.5~TeV$.
By integrating the differential cross section for $500<M_{WW}(GeV)<1500$
and considering an integrated luminosity of 80 $fb^{-1}$ we obtain
127 $W_L$ pairs for the SM and 158 for the BESS model (with $M_V=1~TeV$,
$\gs=13$ and $b=0.01$)
corresponding to a statistical significance of 2.75.
This result is quite discouraging considering the fact that we have not
included the branching ratio.
The situation does not significantly improve by considering a wider
resonance, when varying the BESS parameters in the region allowed by the
present bounds (see Fig. 1).
\par
By increasing the energy to 2 $TeV$ and by considering an integrated luminosity
of 20 $fb^{-1}$, the situation gets even worse.
By integrating the differential cross section for $1000<M_{WW}(GeV)<2000$
we get 7 $W_L$ pairs for the SM and 13 for the BESS model (with $M_V=1.5~TeV$,
$\gs=13$ and $b=0.01$)
corresponding to a statistical significance of 2.27.
\par
The channel corresponding to transverse-longitudinal final $W$'s leads to
a very small bump in the region of the resonance above the SM backgrounds,
which is not observable.
\par
Finally, we have also considered the fusion process in the charged
channel $e^+e^- \rightarrow W^+_{L,T} Z_{L,T}{\overline \nu} e^-$.
However, by comparing with the previous process $e^+e^-
\rightarrow W^+_{L,T} W^-_{L,T}{\overline \nu} \nu$, we can observe that,
in this case, the SM cross section is bigger, being dominated by the
$\gamma W \rightarrow W Z$ fusion process, while the BESS effect
$ W Z \rightarrow V \rightarrow W Z$ is of the same order of magnitude.
So we expect a worse signal to background ratio.
\bigskip
\newcount \nfor

\def \form {\global \advance \nfor by 1 \eqno(6.\the\nfor)}
\noindent
{\bf 6. CONCLUSIONS}
\bigskip
We have shown the usefulness of very energetic $e^+e^-$ linear colliders
in exploring an alternative scheme of electroweak symmetry breaking based on
the existence of vector resonances from strong sector.
\par
Such a study is complementary to those performed in the case of the LHC
(SSC) proton colliders. In fact $pp$ colliders give a valuable opportunity
to study the $V^\pm$ resonances through the $W^\pm Z$ decay [24], while
the $V^0\rightarrow W^+W^-$ channel is difficult to study in $pp$ due to
background problems, and $V^0\rightarrow l^+l^-$ has a very low rate [25].
On the contrary
$e^+e^-$ colliders give the possibility of detecting new neutral vector bosons.
This can be relevant in order to distinguish among the various models with
additional vector bosons. For example, in left-right symmetric models the
charged and the neutral vector resonance masses are split while in the
BESS model they are degenerate (neglecting the electroweak corrections).
\par
Our investigation shows that, even if the mass of the neutral resonance is
higher than the c.m. energy of the collider, the process of $W$
pair production will allow to put very severe restrictions on the parameter
space of the BESS model, especially so if the $W$ polarizations can be
reconstructed from their decay distributions.
\par
If no deviation from the SM prediction is found, already at $\sqrt{ s}=500~GeV$
and integrated luminosity $L=20~fb^{-1}$,
the BESS model parameters $\gs$ and $b$ can be severely
restricted, while for $M_V$ we find significant improvement with respect to
LEP1 when the final $W$ polarization is reconstructed. With higher energy
colliders ($\sqrt{ s}=1~TeV$
and $L=80~fb^{-1}$ and $\sqrt{ s}=2~TeV$
and $L=20~fb^{-1}$) the parameter space contracts and at $b=0$
we get an upper bound on $g/\gs$ of the order of 0.02, for any given value
of $M_V$.
\bigskip
\bigskip
\newcount \nref

\def\ref {\global \advance \nref by 1 \ifnum\nref<10 \item {$ [\the\nref]~$}
\else \item{$[\the\nref]~$} \fi}
\centerline{\bf REFERENCES}

\noindent
\ref 
See "$e^+e^-$ Collisions at 500 $GeV$: the Physics Potential", Proceedings
of the Workshop, edited by P.M. Zerwas, DESY 92, 123A-B, August 1992.
\ref 
See "Physics and Experiments with Linear Colliders", Saariselk\"a,
Finland, September 9-14, 1991, edited by R. Orava, P. Eerola
and M. Nordberd, World Scientific.
\ref 
R. Casalbuoni, P. Chiappetta, S. De Curtis, D. Dominici, F. Feruglio
and R. Gatto, in "$e^+e^-$ Collisions at 500 $GeV$: the Physics Potential",
Proceedings of the Workshop, edited by P.M. Zerwas, DESY 92, 123B,
August 1992, p. 513.
\ref  
D. Dominici, in "Physics and Experiments with Linear Colliders",
Saariselk\"a, Finland, September 9-14, 1991, edited by R. Orava, P. Eerola
and M. Nordberd, World Scientific, p. 509.
\ref  
R. Casalbuoni, S. De Curtis, D. Dominici and R. Gatto, Phys. Lett.
{\bf B155} (1985) 95;
Nucl. Phys. {\bf B282} (1987) 235.
\ref  
M.E. Peskin,
in "Physics and Experiments with Linear Colliders",
Saariselk\"a, Finland, September 9-14, 1991, edited by R. Orava, P. Eerola
and M. Nordberd, World Scientific, p. 1.
\ref 
F. Iddir, A. Le Yaouanc, L. Oliver, O. P\`ene and J.-C. Raynal,
Phys. Rev. {\bf D41} (1990) 22.
\ref  
G.J. Gounaris and J.J. Sakurai, Phys. Rev. Lett. {\bf 21} (1968) 244.
\ref  
J.L. Barklow,
in "Physics and Experiments with Linear Colliders",
Saariselk\"a, Finland, September 9-14, 1991, edited by R. Orava, P. Eerola
and M. Nordberd, World Scientific, p. 423.
\ref 
K. Hikasa
in "Physics and Experiments with Linear Colliders",
Saariselk\"a, Finland, September 9-14, 1991, edited by R. Orava, P. Eerola
and M. Nordberd, World Scientific, p. 451.
\ref  
P. Chiappetta and F. Feruglio, Phys. Lett. {\bf B213} (1988) 95.
\ref 
B. Richter
in "Physics and Experiments with Linear Colliders",
Saariselk\"a, Finland, September 9-14, 1991, edited by R. Orava, P. Eerola
and M. Nordberd, World Scientific, p. 611.
\ref  
I.F. Ginzburg, G.L. Kotkin, V.G. Serbo, S.L. Panfil, and V.I. Telnov,
N.I.M. {\bf 205} (1983) 47.
\ref 
R. Casalbuoni, S. De Curtis, N. Di Bartolomeo,
D. Dominici, F. Feruglio and R. Gatto, Phys. Lett. {\bf B285} (1992) 103;
R. Casalbuoni, A. Deandrea, S. De Curtis, N. Di Bartolomeo,
D. Dominici, F. Feruglio and R. Gatto, UGVA-DPT 1992/07-778, July 1992.
\ref 
L. Rolandi, talk given at the XXVI Int. Conf. on High Energy
Particle Physics, Dallas, Texas, August 1992.
\ref  
CDF Coll., F. Abe et al., Phys. Rev. Lett. {\bf 65} (1990) 2343;
UA2 Coll., J. Alitti et al., Phys. Lett. {\bf B276} (1992) 354.
\ref   
R. Casalbuoni, S. De Curtis, D. Dominici, F. Feruglio and R. Gatto,
Phys. Lett. {\bf B269} (1991) 361.
\ref   
R. Casalbuoni, P. Chiappetta, D. Dominici, F. Feruglio
and R. Gatto, Nucl. Phys. {\bf B310} (1988) 181.
\ref   
A. Djouadi, A. Leike, T. Riemann, D. Schaile, and C. Verzegnassi,
in "$e^+e^-$ Collisions at 500 $GeV$: the Physics Potential", Proceedings
of the Workshop, edited by P.M. Zerwas, DESY 92, 123B, August 1992,
p. 491.
\ref   
K. Fujii, KEK preprint 92-31, to appear in the Proceedings of
the 2nd KEK Topical
Conference on $e^+e^-$ Collision Physics,
KEK, Tsukuba, Japan, November 26-29 1991.
\ref   
J. Layssac, G. Moultaka and F.M. Renard, PM/90-42, December 1990.
\ref   
M. Frank, P. M\"attig, R. Settles and W. Zeuner, in
"$e^+e^-$ Collisions at 500 $GeV$: the Physics Potential", Proceedings
of the Workshop, edited by P.M. Zerwas, DESY 92, 123A, August 1992,
p. 223.
\ref 
M. Kuroda, F.M. Renard and D. Schildknecht, Z. Phys. {\bf C40} (1988) 575.
\ref 
R. Casalbuoni, P. Chiappetta, S. De Curtis, F. Feruglio, R. Gatto,
B. Mele and J. Terron, Phys. Lett. {\bf B249} (1990) 130, and
in "Large Hadron Collider Workshop" Proceedings of the Workshop, edited by
G. Jarlskog and D. Rein, p.786.
\ref 
R. Casalbuoni, P. Chiappetta, M.C. Cousinou, S. De Curtis, F. Feruglio
and  R. Gatto, Phys. Lett. {\bf B253} (1991) 275, and
in "Large Hadron Collider Workshop" Proceedings of the Workshop, edited by
G. Jarlskog and D. Rein, p.731.
\eject
\newcount \nfig

\def\fig {\global \advance \nfig by 1 \ifnum\nfig<10
\item{Fig.  $ \the\nfig~$}\else \item{Fig. $\the\nfig~$} \fi}
\centerline{\bf FIGURE CAPTION}
\noindent
\fig   $90\%$ C.L. contours in the plane $(b,g/\gs)$ for $M_V=1000~GeV$,
       from LEP1 and CDF/UA2 data (we assume
       $\alpha_s=0.12$, and $\Lambda=1000~GeV$) for
       $m_{top}=120~GeV$ (dashed line), $m_{top}=150~GeV$ (solid line),
       and $m_{top}=180~GeV$ (dotted line).
       The allowed regions are the internal ones. For increasing $\alpha_s$
       the allowed regions get shifted to the left.

\fig   $90\%$ C.L. contours in the plane $(b,g/\gs)$ for $\sqrt s=500~GeV$
        and $M_V=600~GeV$
       from the fermion channel (we assume $m_{top}=150~GeV$,
       $\alpha_s=0.12$ , $\Lambda=1000~GeV$). The solid line corresponds to
        polarization $P_e=0.5$ while the dashed line is for unpolarized
       electron beams.
       The allowed regions are  the internal ones.

\fig   $90\%$ C.L. contours in the plane $(M_V,g/\gs)$ for $b=0$,
       from the fermion channel (we assume $m_{top}=150~GeV$,
       $\alpha_s=0.12$, $\Lambda=1000~GeV$, and polarization $P_e=0.5$)
       for different choices of $\sqrt s(GeV)$: 500 (solid line),
       1000 (dashed line).
       The lines give the upper bounds on $g/\gs$.

\fig   Unpolarized and longitudinally polarized differential cross-section
       $d\sigma/d\cos\theta$
       (in $pb$) in the $WW$ channel, for one $W$ decaying leptonically and
        the other hadronically, versus $\cos\theta$
       for the SM (dash line) and BESS model (solid line)
       corresponding to $\sqrt s =1000~GeV$, $M_V=1500~GeV$, $b=0$, and
       $g/\gs=0.05$.
       The error bars are the total errors and correspond to one standard
       deviation. The lower curves are for the $W_LW_L$ channel.

\fig   Same of Fig. 4 for $\sqrt s=2000~GeV$
        and $M_V=2500~GeV$.

\fig   $90\%$ C.L. contours in the plane $(b,g/\gs)$ for $\sqrt s =500~GeV$
       and $M_V=600~GeV$ from the unpolarized
       $WW$ differential cross section (dotted line), from  the
       $W_{L}W_{L}$ differential cross section
       (dashed line) and  from all the
       differential cross sections for $W_{L}W_{L}$, $W_{T}W_{L}$,
       $W_{T}W_{T}$ combined with the $WW$ left-right
       asymmetries (solid line). The allowed regions are the internal ones.

\fig   $90\%$ C.L. contour in the plane $(b,g/\gs)$ for $\sqrt s =500~GeV$
       and $M_V=600~GeV$ from all the
       differential cross sections for $W_{L}W_{L}$, $W_{T}W_{L}$,
       $W_{T}W_{T}$ combined with the $WW$ left-right
       asymmetries, with the total number of events reduced by a factor
       0.5 to account for possible losses
       due to the reconstruction of the polarization
        of the $W$'s. The allowed region is the internal ones.

\fig   $90\%$ C.L. contour in the plane $(b,g/\gs)$ for $\sqrt s =2000~GeV$,
       an integrated luminosity of 20 $fb^{-1}$
       and $M_V=2500~GeV$ from the unpolarized
         $WW$ differential cross section.
       The allowed region is the internal ones.

\fig   $90\%$ C.L. contours in the plane $(M_V,g/\gs)$ for $\sqrt s=500~GeV$,
       $L=20~fb^{-1}$ and $b=0$.
       The solid line corresponds to the bound from
       the unpolarized $WW$ differential cross section,
       the dashed line to the bound
       from the longitudinally polarized
       $W_{L}W_{L}$ differential cross section,
       the dotted line to the bound from all the polarized
         differential cross sections $W_{L}W_{L}$, $W_{T}W_{L}$,
      $W_{T}W_{T}$   combined with
       the $WW$ left-right asymmetries.
       The lines give the upper bounds on $g/\gs$.

\fig   Same as Fig. 9 for $\sqrt s=1000~GeV$ and $L=80~fb^{-1}$.
       Here the dashed and the dotted lines are almost coincident.

\fig   Same as Fig. 10 for $\sqrt s=2000~GeV$ and $L=20~fb^{-1}$.
       Here the dashed and the dotted lines are coincident.

\fig   Longitudinally polarized differential cross-section
       $d\sigma/d M_{WW} (e^+ e^- \to W^+_L W^-_L \nu\bar\nu)$
       (in $fb/GeV$)  versus $M_{WW}$
       for the SM (dash line) and BESS model (solid line)
       corresponding to $\sqrt s =1500~GeV$, $M_V=1000~GeV$, $b=0.01$, and
       $\gs=13$.

\fig  Same of Fig. 12 for $\sqrt s =2000~GeV$, $M_V=1500~GeV$, $b=0.01$, and
       $\gs=13$.
\bye